\documentclass{optica-article}

\journal{opticajournal} 

\articletype{Research Article}

\usepackage[overlay,absolute]{textpos}\newcommand\PlaceText[3]{\begin{textblock*}{10in}(#1,#2)#3\end{textblock*}}

\begin{document}

\PlaceText{12mm}{8mm}{Optica Quantum \textbf{2}, 5/25 (2024); https://doi.org/10.1364/OPTICAQ.534258}

\title{Metro-scale QKD Using Multimode Fiber}

\author{A. Brzosko,\authormark{1,2,*} R. I. Woodward, \authormark{1} Y. S. Lo,\authormark{1} M. Pittaluga,\authormark{1} P. R. Smith,\authormark{1} J. F. Dynes \authormark{1} and A. J. Shields\authormark{1}}

\address{\authormark{1}Toshiba Europe Limited, 208 Cambridge Science Park, Cambridge, CB4 0GZ, UK}
\address{\authormark{2}Cambridge University Engineering Department, 9JJ Thomson Avenue,
Cambridge, CB3 0FA, UK}

\email{\authormark{*}ab2967@cam.ac.uk} 

\begin{abstract*} 
We report a proof-of-principle realisation of a decoy-state BB84 QKD protocol with phase encoding over a record-breaking $17$ km of MMF at a rate of $193$ kbits/s, as well as over $1$ Mbit/s at a distance of $1$ km. These results suggest that QKD can be deployed over MMF in metropolitan-scale telecommunication connections. Such MMF metropolitan networks are ubiquitous - thus this advance could pave the way to wide scale metropolitan deployment.
We also assess the advantages of adapting the OM3 channel using mode-matching photonic lanterns on the QBER, signal gain, and key rate and compare different encoding techniques in light of MMF propagation effects. 
This work confirms the suitability of current QKD technology for use in existing MMF links, unlocking new opportunities for quantum applications using legacy fibre. 

\end{abstract*}

\section{Introduction}

\begin{figure}[ht!]
\centering\includegraphics[clip,trim={0 1cm 0 1cm },width=0.8\textwidth]{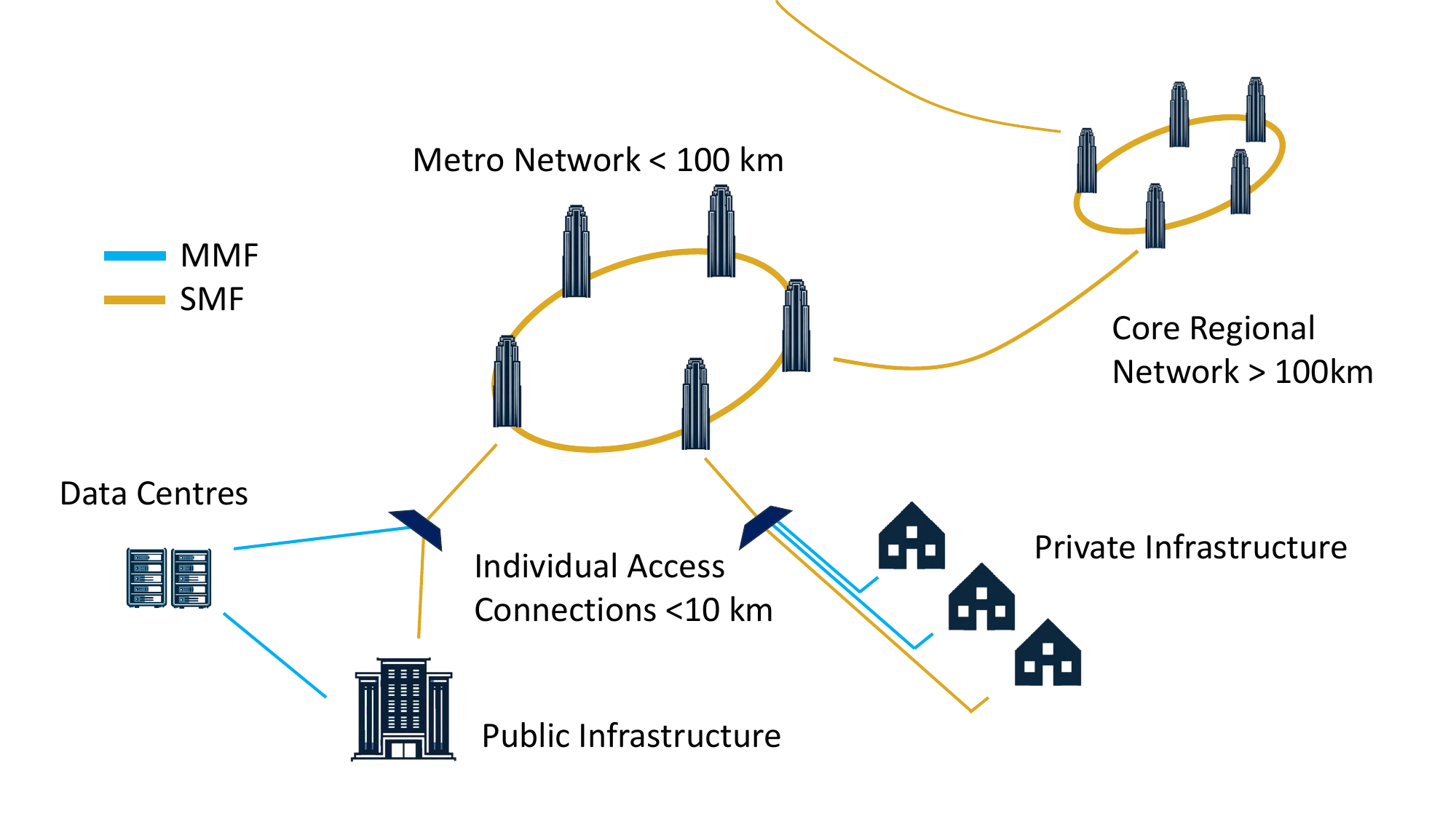}
\caption{A typical layout of an optical network at different scales. The shortest links may be historic MMF (OM1/2) or newly deployed (OM3/4/5), connecting data centers, public utility buildings, company buildings or households. They are highlighted in light blue. Longer links are SMF.}
\label{Img:Map}
\end{figure}

The first optical telecommunication networks utilized fibers with core diameters on the order of tens of micrometers, which are typically multimodal in the telecom wavelength ranges. Although the long- and medium-haul fiber network infrastructure now uses the single-mode standard, many metro-scale connections remain as multimode fibre (MMF). This is mainly due to the fact that modern postprocessing techniques allow the resolving of intermodal (and indeed intramodal) dispersive effects at multi-kilometer distances for classical communications \cite{Downie.2011}. Moreover, replacing existing connections with higher quality fiber is expensive and can rarely be justified for low bandwidth requirement links. Often the existing legacy connections are left in place even if new fiber is laid. Such connections may now be dark, as the active channels have been moved over to the higher performance fibers. These dark connections could present a promising opportunity for emerging quantum technologies if the MMF could be harnessed for transporting quantum signals, as the lack of bright telecommunications channels reduces the noise from related effects, such as Raman scattering \cite{Patel.2012}. An example composition of a modern optical network at various hierarchical levels is shown in Fig. \ref{Img:Map}, with MMF links highlighted. 

Using MMF for quantum signal transmission is not without its challenges however. There are issues with both launching light into the fiber and transmission itself. There is normally a mismatch between the mode field diameter (MFD) of different types of fibre - here SMF (single mode fiber) and MMF - which causes losses and signal distortion. Moreover, depending on the properties of the profile of the wave front entering the fiber, the light may propagate in different modes. These not only have different propagation coefficients, which affect the speed of transmission, loss and phase evolution, but also couple to each other, causing the signal to interfere with itself and with neighboring signals. This normally leads to additional constraints on the bandwidth (due to modal dispersion), and for longer distances can prohibit phase-encoded communication by scrambling the wave front. Furthermore, MMF has nominally higher attenuation coefficient than SMF, which can limit the range of optical communication.

Quantum Key Distribution (QKD) is a physically secure method for distribution of cryptographic keys relying on principles of quantum mechanics \cite{Bennett.2014, Ekert.1991, Pirandola.2020}. Such keys can be used for encryption in symmetric key cryptography protocols, like AES, or for information-theoretically secure encryption, e.g. a one-time pad. Since its inception, the field of QKD has progressed greatly, eliminating many implementation imperfections \cite{Mayers.1998, Lo.2012} and achieving greater key generation rates and distances \cite{Lucamarini.2018}. Recently, theoretical advancements have been made in quantum information processing within MMF, such as new computation and communication protocols and devices \cite{Xavier.2020, Amitonova.2020}. However, a knowledge gap still exists in the execution of high performance QKD over MMF.

\begin{figure}[ht!]
\centering\includegraphics[clip,trim={0 3cm 0 2cm },width=1\textwidth]{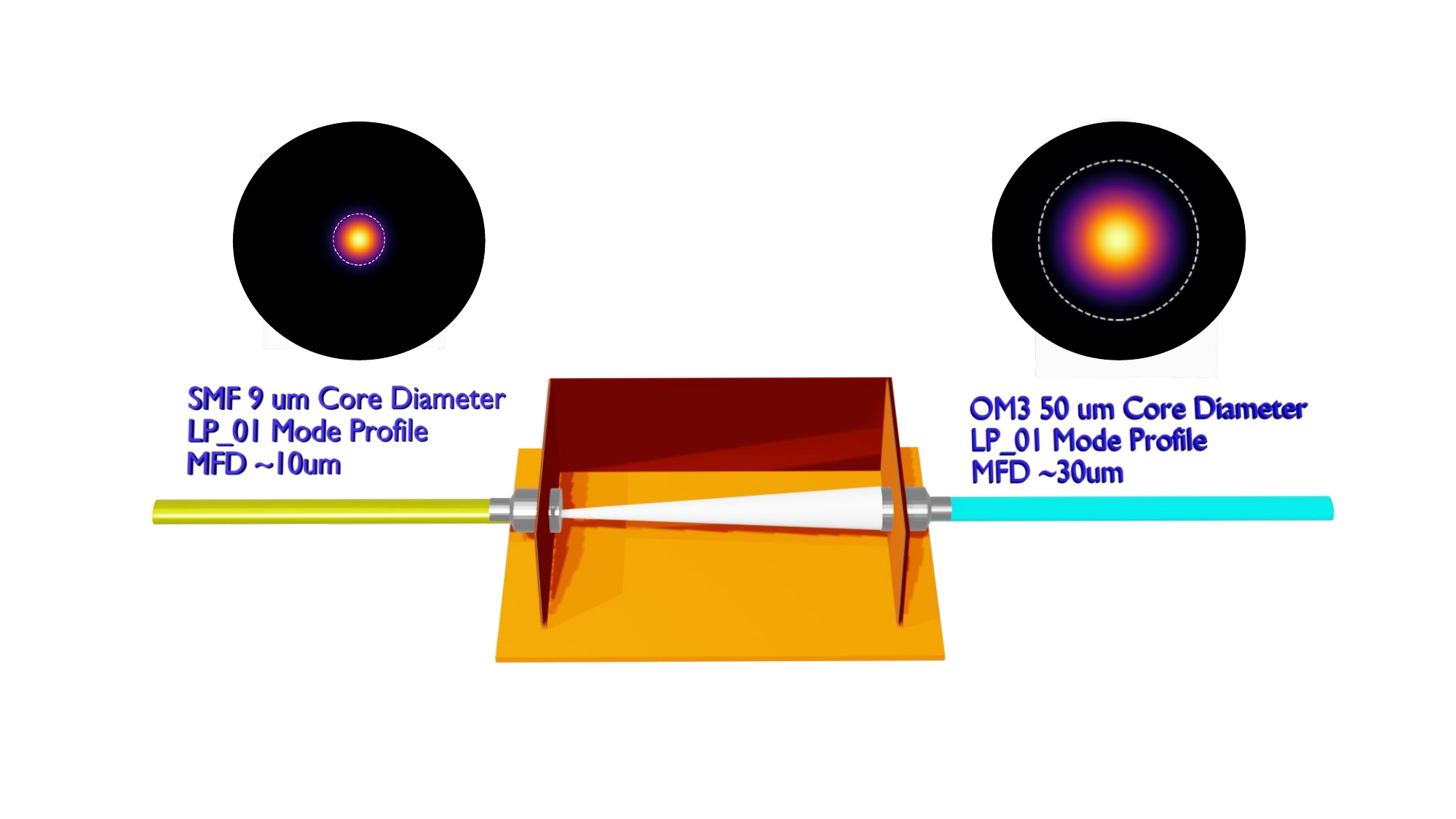}
\caption{The conceptual schematic of the mode-field adapter used in our experiment, showing the tapered waveguide mapping the LP01 mode field distribution from SMF onto OM3. Mode field intensity plots are shown as inserts, with the dotted white circle indicating the fibre core.}
\label{Img:adapter}
\end{figure}

The topic of QKD using a MMF communication channel is underexplored and indeed, the only prior demonstration  of this achieved 137 bit/s over 550 m of installed graded-index MMF in 2005\cite{Namekata.2005}. In this paper we demonstrate a significant advance in this area by experimentally realising a proof-of-principle BB84 protocol with phase-encoding and biased bases over multiple kilometers of the most popular OM3 (Optical Multimode 3rd generation) fibre standard, carry out an asymptotic key analysis of a 2-decoy variant, and discuss practical aspects affecting the use of MMF for quantum technologies. Furthermore, we present an improvement to the performance that can be achieved with a mode-matching photonic lantern device, which is a tapered waveguide adapter matching the intensity distributions from MMF to SMF and vice versa \cite{Gross.2014}. The action of the mode-matching adapter is shown conceptually in Fig. \ref{Img:adapter}.

\section{Experiment and Results}

\begin{figure}[ht!]
\centering\includegraphics[clip,trim={0 3cm 0 3cm },width=1\textwidth]{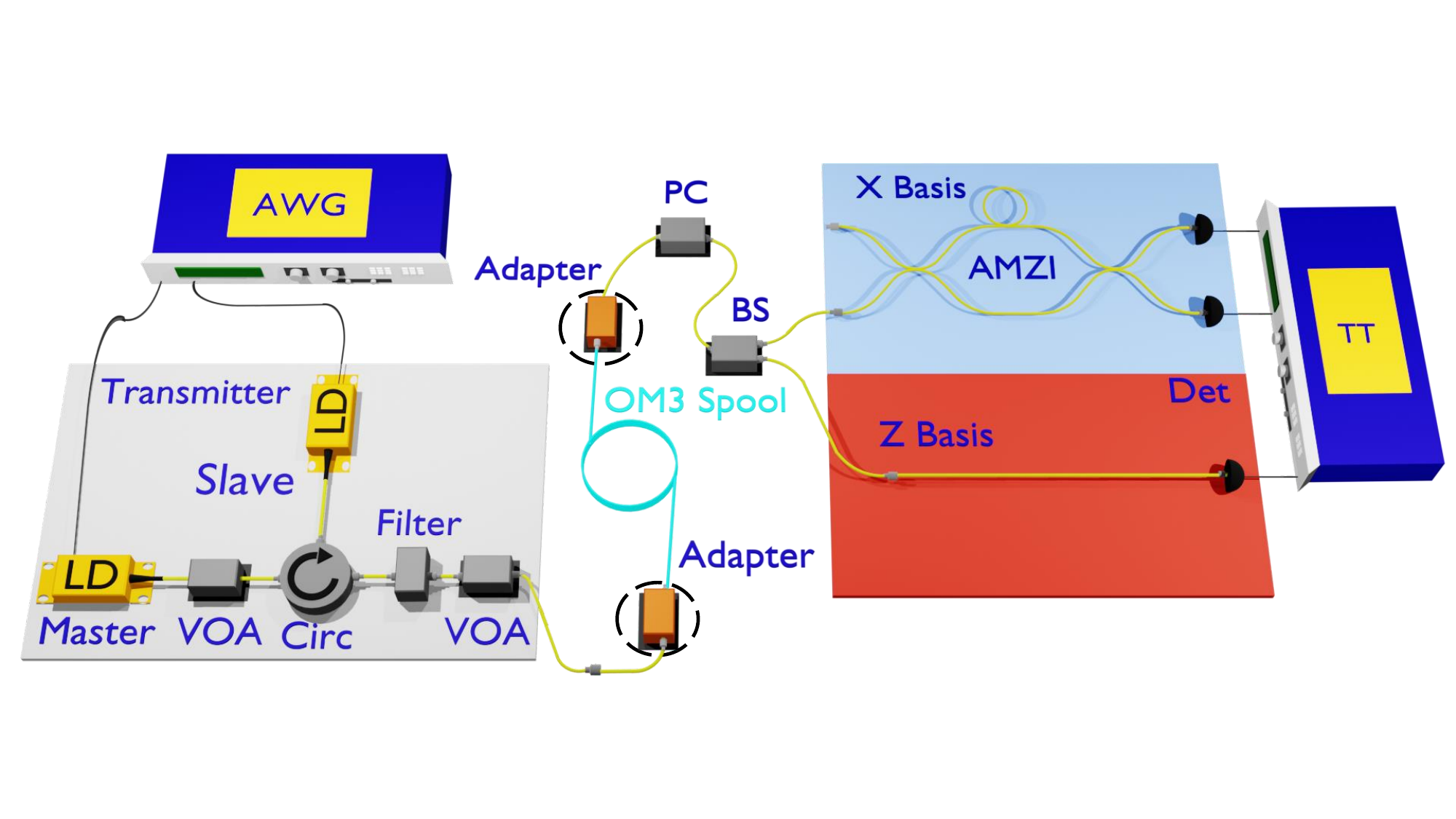}
\caption{Experimental setup. The transmitter has Laser Diodes (LDs) receiving electrical pulses from the Arbitrary Waveform Generator (AWG). The receiver is either in the X basis (light blue) or Z basis configuration (light red). The adapters (in black circles) are removed for some measurements. Circ: circulator, PC: polarisation controller, BS: beam splitter, VOA: variable optical attenuator, Det: detectors, TT: timetagger.}
\label{Img:Setup}
\end{figure}

Our experimental setup is shown in Fig. \ref{Img:Setup}. We developed a QKD transmitter, able to encode information into the relative phase (X basis) and time of arrival (Z basis) of weak coherent light pulses. The transmitter is composed of two Distributed Feedback Laser Diodes (LDs) operating at 1550 nm, with one (slave) optically injection-locked \cite{Paraiso.2021} to the other (master) via a circulator. The system adopts an all-optical direct phase modulation scheme, as introduced in \cite{Yuan.2016}. The light from the master laser, pulsed at 1 GHz, is used to phase-seed the pulses in the slave laser cavity, which is pulsing at 2 GHz. The phase of each master laser pulse is random, since there is enough time between the consecutive pulses for the phase coherence in the cavity to diffuse. However, the pairs of slave pulses seeded by the same master pulse will be in-phase with each other. This means that when we delay a slave pulse and interfere it with the following one from the same pair on a beam splitter, they will exit out of the same output port. We can then encode a complementary digit by changing the phase between these pairs of pulses by $\pi$, so that they exit out of the other output port when interfered. This is done by directly modulating the master laser electrical driving pulse. The time encoding in the Z basis is simply an early or late pulse of the secondary laser being switched off within the time window of a primary laser pulse. 

The pulses produced are spectrally filtered to 0.1 nm width to reduce the amplified spontaneous emission noise, and then attenuated to single-photon level (mean photon number $\mu=0.4$ photons/pulse) before being sent down an MMF channel. The decoder for the phase encoding is an asymmetric Mach-Zehnder interferometer (AMZI) with a 500 ps delay in one of the arms. The AMZI output is monitored with a superconducting nanowire single photon detector (SNSPD). The time-bin encoded pulses are simply detected with the SNSPD. 

Both the transmitter and receiver use SMF. Even when targeting MMF communication channels, light source and detector technology will still primarily comprise of SMF fibre-based components due to the wide availability of such equipment off-the-shelf. The transmitter and receiver are linked with a MMF channel, which is composed of spools of graded-index OM3 fiber supporting 8 spatial mode groups at 1550 nm \cite{Franz.2013}. The lengths used are 1, 2, 5 and 10 km, as well as their combinations, interconnected using MMF patch cables to achieve distances of 3, 7, 8, 12, 15 and 17 km. For means of comparison, for every distance we take all measurements in two configurations: once with the underfill launch and once with the mode-matching adapters at both ends of the channel. These adapters adiabatically match the SMF mode profile with the fundamental mode of the MMF and suppress the coupling into higher order modes \cite{Gross.2014}. For the underfill launch configuration, the channel is connected directly to the SMF output of the transmitter and the SMF input of the receiver. This ensures the launch into an underfilled mode distribution, since the MFD of the LP$_{01}$ mode of the SMF (the only mode) is $\sim$10 \textmu m, while the MFD of the LP$_{01}$ of the MMF, which constitutes a mode group on its own, is $\sim$30 \textmu m \cite{Yu.2021,Kaczmarek.2004}. The receiver attenuation is 3 dB. The efficiency of the SNSPD is 50\%, with a dark count rate of less than 10 Hz. 

\begin{figure}[ht!]
    \centering
    \includegraphics[width=1\textwidth]{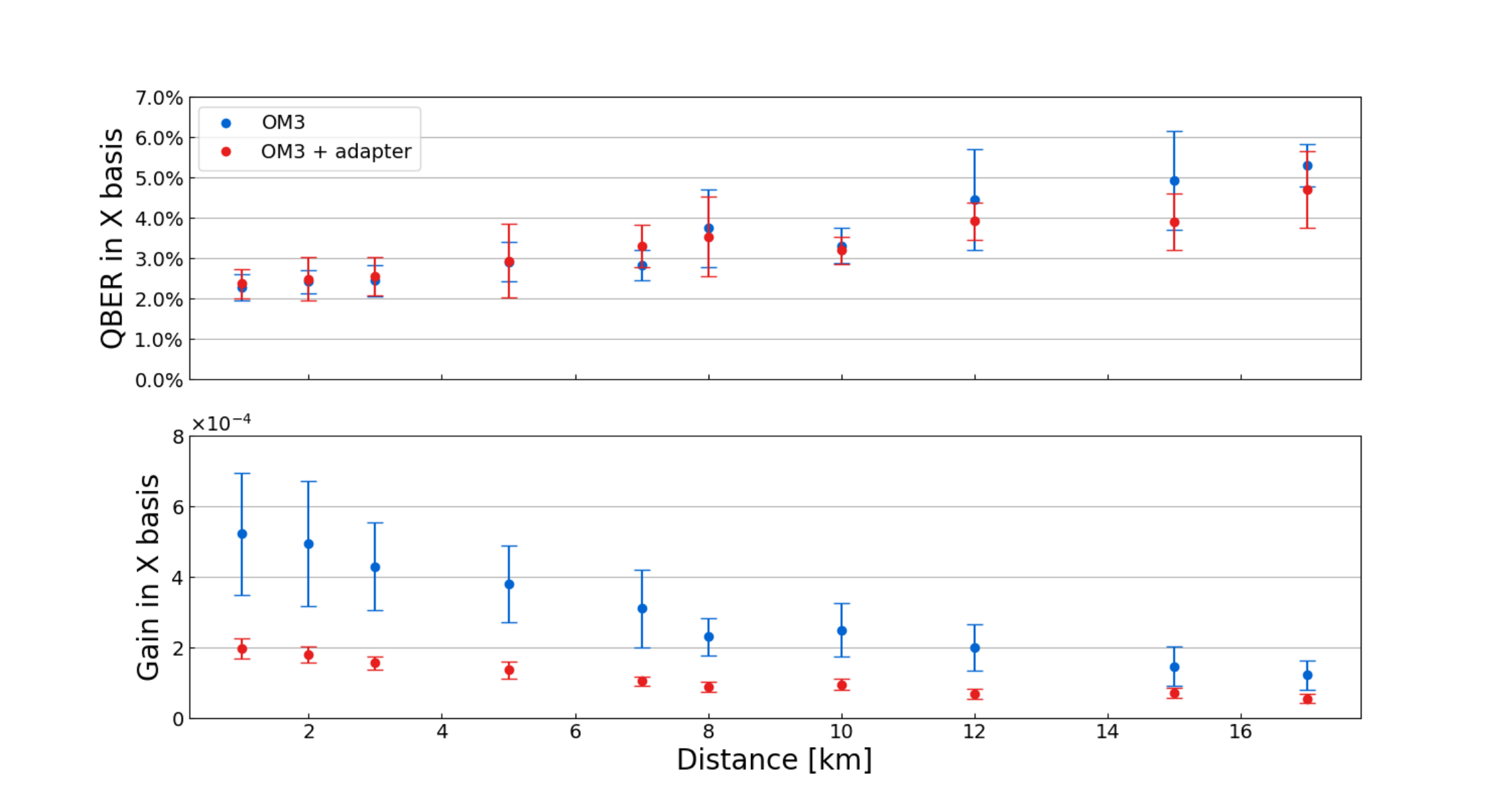}
    \includegraphics[width=1\textwidth]{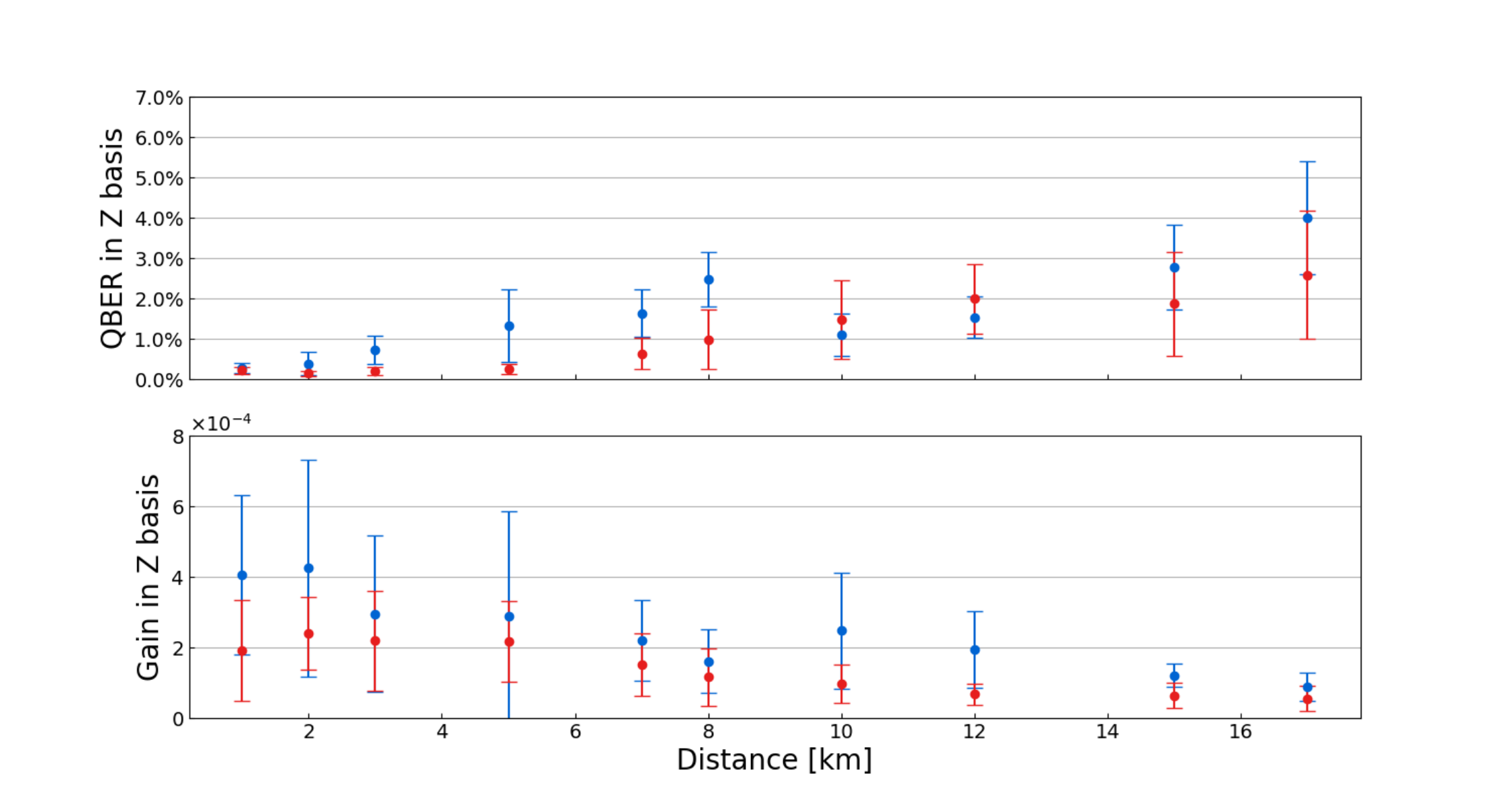}
    \caption{QBER and gain values for each distance with (red) and without (blue) adapters for both X and Z bases. The error bars show the standard deviation of the mean from 5 repetitions of the experiment.
    }
    \label{Img:QBERsGains}
\end{figure}

For each basis and spool length, we repeatedly send a fixed random train of 1000 bits and time tag the SNSPD event detections from which a histogram of events over a 10 second acquisition period is built. We then calculate the quantum bit error rate (QBER) and signal gain by aligning the received and sent patterns and comparing them. The QBER is calculated by gating each pulse on the histogram to its central 250 ps and taking the ratio of photons in the incorrect pulses to the sum of all the photons in the gated pulses. The gain is the ratio of the photons registered by the SNSPD to the ones sent (0.4 photons per qubit at 1 GHz over 10 s). This process is carried out with and without the mode-matching adapters. One of the challenges using MMF is the potential for light to couple between modes during propagation, which is enhanced by variation in fibre properties due to thermal or mechanical changes caused by the external environment. To include these effects, we perform the experiment 5 times for each distance on different days and average the results. These are shown in Fig. \ref{Img:QBERsGains}. The statistical errors in QBER tend to increase with distance, while the opposite is true for the gain. This shows that the fractional error is the same for all distances, that is the errors at each distance are proportional to signal strength. For example, in the gain of X basis, the error at 5 km is about $\pm1 \cdot 10^{-4}$, which corresponds to about 25\% of the mean value, while at 12 km it is $\pm0.5 \cdot 10^{-4}$, which again is about 25\% of the mean value. In both bases the gains are similar, while the QBERs are noticeably lower in the Z basis. There is a slight improvement in QBER and decrease in gain when using adapters.

\begin{figure}[ht!]
    \centering
    \includegraphics[width=1\linewidth]{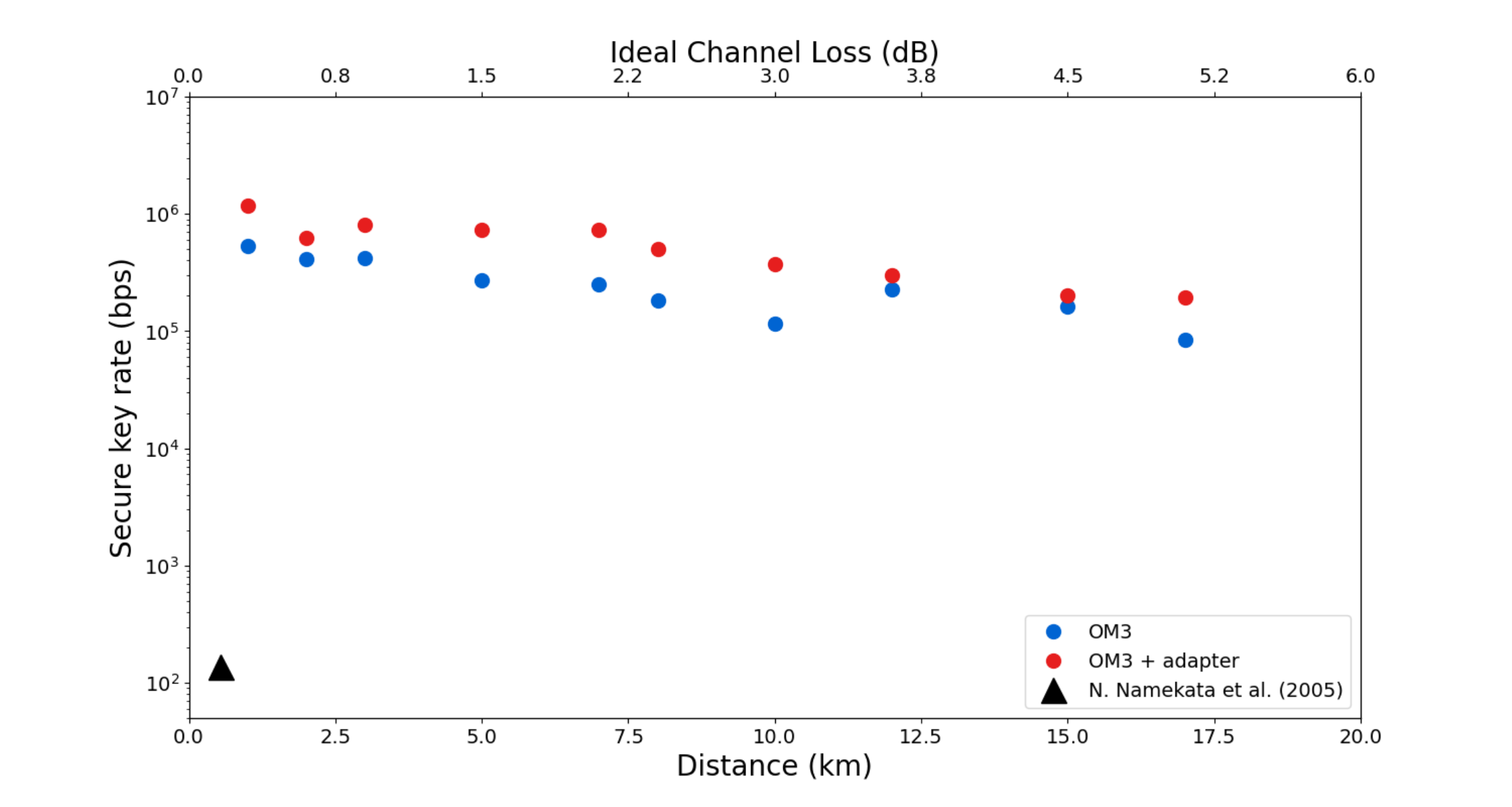}
    \caption{Estimated SKR for each measured channel loss with (red) and without (blue) the mode-matching adapters. The ideal channel loss is calculated by assuming 0.3 dB/km in OM3.
    }
    \label{Img:SKR}
\end{figure}

Finally, we use the two measured quantities (QBER and gain) to estimate the Secure Key Rate (SKR) for each distance, using the 2-decoy asymptotic key analysis from \cite{Ma.2005} and \cite{Lo.2005}, including a consideration of ratios of basis choices and decoy state probabilities from \cite{Lucamarini.2013}. We also characterize the loss for each spool length, as these differ from the losses expected from the quoted loss per distance for each spool, due to the impact of launch conditions, as described in more detail in the next section. 
The result is shown in Fig. \ref{Img:SKR}. The equivalent channel loss is that quoted per distance, not the measured value. We achieve $1.18$ Mbit/s SKR at 1 km with the adapters (0.54 Mbit/s without), which can be contrasted with 137 bit/s over 550 m from \cite{Namekata.2005}. It is worth noting here, that the previous demonstration was performed with InGaAs detectors at 100 kHz clock rate. The significantly improved performance we demonstrate here is due to an enhanced system design which enables us to operate at 1 GHz clock speed, as well as using more efficient single photon detectors. The estimated SKR decreases with distance, reaching 193 kbit/s (84 kbit/s without the adapters) at the maximum measured distance of 17 km. Despite the adapters decreasing the signal gain, their combined impact on QBERs in both bases is enough to improve the SKR at all distances by up to 3 dB.

\begin{figure}[ht!]
    \centering
    \includegraphics[width=0.9\textwidth]{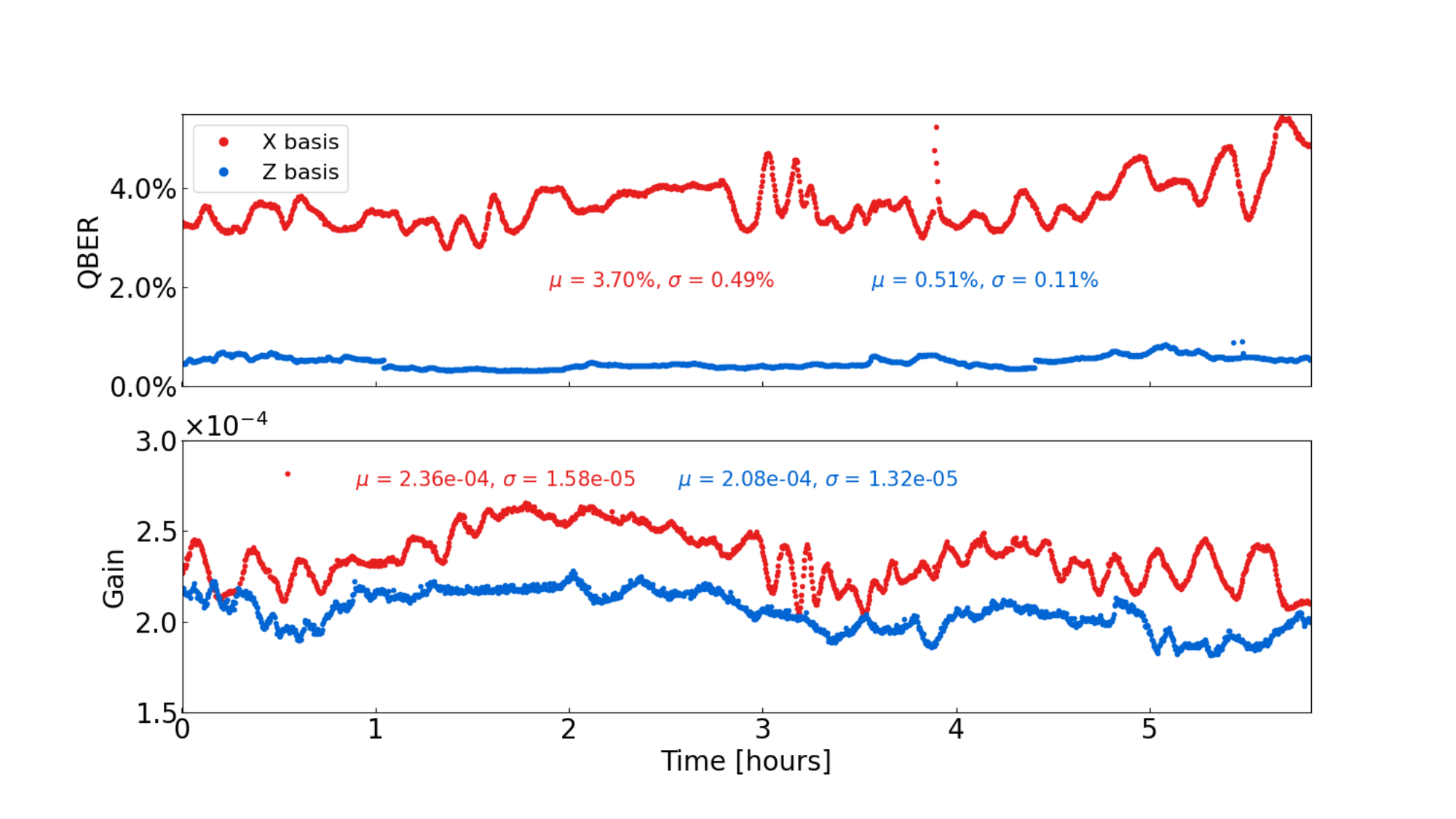}
    \caption{QBER (red) and gain (blue) parameters measured every 10 seconds for X and Z basis encoding without the mode-matching adapters at a distance of 10 km of OM3 over a period of almost 6 hours. The gain variation is similar for both bases, while for the X bases the QBER not only is worse but also varies more strongly than for the Z basis.}
    \label{Img:Stability}
\end{figure}

Following the estimation of the SKR at different distances, we investigated its stability over time under the impact of normal temperature variation and vibrations within a laboratory setting. We measured the QBER and gain every 10 seconds for each distance and encoding basis over multiple hours. One of these measurements at 10 km is shown in Fig. \ref{Img:Stability}. Both metrics oscillate on the scales of minutes, with changing amplitude of oscillation. The period of these oscillations aligns well with the air conditioning cycle in the lab.

The phase (X) basis not only has a higher QBER than Z basis, but is also less stable. Apart from the obvious difference in QBER fluctuations between the bases, this can be deduced from the fact that the gain in both bases has a similar variation.

In both bases, there is a correlation between QBER and gain on the scale of minutes. Again, this phenomenon is easier to note in the X basis, where the fluctuations are more pronounced. It can be attributed to the temperature variation affecting the amount of intra-mode coupling due to thermally induced mechanical expansion that subtly alters the fibre geometry / launch condition at connectors, which changes both the loss and the number of phase and timing errors. In general more intra-mode coupling will increase the loss as well as the difference in the phase evolution speed within qubits, therefore causing more phase errors. The timing errors will result from the different propagation speeds in different modes. These effects may only be quantified by the proxy metrics of variation in QBER and gain. It is noted that longer MMF propagation distances result in greater time/phase delays between modes, effecting in greater QBERs as shown in Fig \ref{Img:QBERsGains}.

In practice, the QBER and gain variation will affect the key generation rate, but in our experiment it was never significant enough to prevent the protocol from execution. For example, in Fig. \ref{Img:Stability} the standard deviation of the QBER was 0.49 percentage points for X basis and 0.11 percentage points for the Z basis (equivalent to a relative fluctuation of 13\% and 22\% respectively), while the gain varied by only about 6.5\% for both bases. Moreover, studies to date have shown that phase drift affects deployed fiber less than that in a laboratory setting \cite{Liu.2021}, which bodes well for practical use. 

\section{Discussion}
In our experiment, we estimated the SKR at different distances of OM3 via QBER and gain measurements using two different launch methods into MMF - the direct underfill launch and an adiabatically matched one. Both approaches enabled us to obtain scaling equivalent to SMF with a 0.3 dB/km loss. 

Our results, which differ from typical SKR estimates over SMF that generally follow a simulated trend line, vary chaotically - with the expected behaviour difficult to infer.
This comes from the richness of MMF physics, most prominently the sensitivity of mode of propagation to the launch conditions and the coupling between different modes during propagation \cite{Franz.2013}. Any instance of the light travelling in separate modes - which is unavoidable - will introduce both errors and loss in the communication. This is due to each mode having different propagation constants and loss coefficients, which leads to spatial dispersion of the signal as well as difference in phase evolution speed in different modes. Moreover, the modal distribution depends on the temperature fluctuations, as well as any macrobends, as they affect the components of the stresses on the fiber structure and therefore its optical properties \cite{Boechat.1991, Alarabi.2018}.

The underfilled mode distribution was achieved by directly connecting SMF and MMF with LC connectors. However, the MMF connectors and ferules exhibit coarse manufacturing precision with respect to  the concentricity, ellipticity and diameter uncertainty, which were not characterized. Due to this, the launch conditions - the angle and the gap between the connector ferules - are slightly different every time we make a connection. Therefore, each measured value of gain and QBER is effectively sampled from a distribution of possible sub-ideal values and leads to deviation from an average expected behavior. For this reason, we have taken five measurements for each distance and encoding basis and used the statistical means of these for SKR estimation. 

In the same manner we repeated the measurements using the tapered mode-matching adapters. Here, the light distribution is mapped from the SMF onto the fundamental mode (LP$_{01}$) of the MMF via a widening taper. After travelling in the channel, a reverse procedure is applied to the signals. This reduces the proportion of energy launched into higher order modes and subsequently minimizes coupling between different modes within MMF. This effect can be seen in the resulting estimates for SKR, which are better than the underfill launch method. In the literature, we observe that there is increasing interest in using MMF for satellite QKD transmitter/receiver systems, to enable greater collection efficiency \cite{Lee.2023} - our results introducing mode tailoring using such adapters could thus have useful applications to the free-space QKD domain as well.

The variable launch conditions and change of intra-mode coupling with temperature also lead to an inconsistency in loss variation with distance. This introduces additional smearing effect to the spread of our key rates at each distance. Despite well-defined spool lengths in our experiment, the measured losses have a more noisy distribution, with an additional non-constant difference between the data with and without adapters due to different connectors. 

Many QKD systems use two phase bases for encoding - we also use it for one of our basis for means of comparison. We chose the majority basis to be time-bin encoding, as it has a lower QBER characteristic than phase, therefore improving SKR. It is also more stable against MMF fibre fluctuations than the phase basis, since the phase is scrambled by interference between different mode groups with differing phase evolution speeds in addition to suffering from the dipsersive effects.

In terms of practical QKD deployments, consideration should also be given to the transport of the classical service channel that is required for a full QKD protocol. While our proof-of-principle experiment employed out-of-band communication between Alice and Bob, it is still possible to copropagate quantum and classical signals with an MMF channel. For example, there are already existing demonstrations of wavelength-division in OM3 fiber \cite{Shukla.2023} and other generations of the OM standard \cite{Benyaha.2018} over kilometer-scale distances. In our setup, such multiplexer would simply be inserted after the transmitter but before the SMF-MMF connection and analogously a demultiplexer after the MMF-SMF connection but before the receiver and would be compatible with both launch methods. Additionally, multiplexing quantum and classical signals has also been demonstrated in spans of conventional SMF at shorter wavelengths (850 nm with 1300nm and 850 nm with 1550 nm) at which it can support multiple modes \cite{Gordon.2004,Ramos.2024}. Finally, there may not be a need to co-propagate a classical channel along the quantum channel in MMF, if it links two locations which are connected to the internet - the reconciliation can be then done over some other public channel. This is particularly applicable for internet-enabled data centres connected with dark MMF.

Our work paves the way to other exciting investigations of QKD over MMF. It is interesting to determine whether different mode groups can be selected as channels for the quantum signal and how this choice affects the communication performance. Following from there, we should establish if we can carry out the multiplexing of multiple quantum and classical signals in different spatial modes. Additionally, we could use different spatial modes as an encoding basis, which possibly allows encoding qudits in dimension larger than 2. Such concepts have recently been demonstrated using multi-core fibre (MCF) \cite{Dynes.2016, Zahidy.2024} but it remains an open question whether the approaches can perform well using MMF, which is already much more widely deployed than MCF. This is very relevant to the current research direction in fiber-based communications, as we are quickly approaching the (nonlinear) Shannon limit in possible dimensions of channel multiplexing (wavelength and time) and all degrees of freedom of the optical field (polarization, phase and amplitude). Specifically, there is a rising interest in exploring the spatial dimension for information transfer \cite{Rademacher.2021}.

\section{Conclusion}

We have demonstrated a proof-of-principle realisation of the BB84 protocol in OM3 fibre. Using the phase and time bin encodings, we achieved QBER and gain results, which would allow for secure key generation rates exceeding 1 Mbit/s at practical distances for MMF deployment. We have also shown that a key can be obtained at a distance of 17 km of OM3 at an estimated rate of 193 kbit/s, which is the longest distance at which QKD has been demonstrated in MMF to the best knowledge of the authors. Our experiment involved a standard single-mode-fibre based QKD transmitter and receiver with standard connectors between both MMF and SMF. Although the use of MMF results in worse performance than an equivalent length of SMF, the achieved key rates are still suitable for practical use, thus unlocking the prospect of using the large quantity of already deployed MMF fibre for quantum applications. Additionally, we have demonstrated that using a plug-and-play mode-matching adapters can directly improve the key generation rate in our experiment.

\begin{backmatter}
\bmsection{Funding}
A.B. gratefully acknowledges financial support from the EPSRC ICASE (Award No. 210138) and Toshiba Europe Limited

\bmsection{Author contributions} A.B. developed the experimental system, performed the experiment, and analysed the data. R.I.W and M.P. provided the idea for the demonstration and guidance throughout. Y.S.L. helped with assembling the transmitter and receivers and writing the analysis code. P.R.S. and J.D. helped with smooth running of the experiment. A.S. guided the work. A.B. wrote the manuscript, with contributions from all the authors.

\bmsection{Acknowledgments}
The authors acknowledge helpful discussions with Professor Richard Penty.

\bmsection{Disclosures}
\medskip
\noindent The authors declare no conflicts of interest.

\bmsection{Data availability} Data underlying the results presented in this paper are not publicly available at this time but may be obtained from the authors upon reasonable request.

\end{backmatter}
\bibliography{bibliography}

\end{document}